\documentclass[12pt,preprint]{aastex}

\shorttitle{Flux rope} \shortauthors{Li & Zhang}

\begin{document}

\title{Homologous Flux Ropes Observed by SDO/AIA}

\author{Ting Li\altaffilmark{1} \& Jun Zhang\altaffilmark{1}}

\altaffiltext{1}{Key Laboratory of Solar Activity, National
Astronomical Observatories, Chinese Academy of Sciences, Beijing
100012, China; [liting;zjun]@nao.cas.cn}

\begin{abstract}

We firstly present the \emph{Solar Dynamics Observatory}
observations of four homologous flux ropes in active region (AR)
11745 on 2013 May 20-22. The four flux ropes are all above the
neutral line of the AR, with endpoints anchoring at the same region,
and have the generally similar morphology. For the first three flux
ropes, they rose up with a velocity of less than 30 km s$^{-1}$
after their appearances, and subsequently their intensities at 131
{\AA} decreased and the flux ropes became obscure. The fourth flux
rope erupted ultimately with a speed of about 130 km s$^{-1}$ and
formed a coronal mass ejection. The associated filament showed an
obvious anti-clockwise twist motion at the initial stage, and the
twist was estimated at 4$\pi$. This indicates that kink instability
possibly triggers the early rise of the fourth flux rope. The
activated filament material was spatially within the flux rope and
they showed consistent evolution in their early stages. Our findings
provide new clues for understanding the characteristics of flux
ropes. Firstly, there are multiple flux ropes that are successively
formed at the same location during an AR evolution process.
Secondly, a slow-rise flux rope does not necessarily result in a
CME, and a fast-eruption flux rope results in a CME.

\end{abstract}


\section{Introduction}

The definition of homologous coronal mass ejections (CMEs) was
originally clarified  by Zhang \& Wang (2002). Homologous CMEs must
originate from the same region, have a similar coronagraphic
appearance, be associated with homologous flares and similar EUV
(X-ray) dimmings. The study of homologous CMEs is of great
importance to understand the triggering mechanisms of CMEs. Emerging
magnetic flux and photospheric flows including shearing motions and
converging motions were considered to be the initial causes of
homologous CMEs (Nitta \& Hudson 2001; Chertok et al. 2004; DeVore
\& Antiochos 2008; Soenen et al. 2009; Wang et al. 2013).

The flux rope is thought to be closely related to the CME and almost
all theoretical models of CMEs require the presence or formation of
a coronal magnetic flux rope (Forbes 2000). CMEs generally have a
three$-$part structure: the bright core, the dark cavity and the
leading edge (see e.g., Illing \& Hundhausen 1986). There exists a
viewpoint that the twisted flux rope corresponds to the dark cavity
which accumulates magnetic energy and mass within it (Hudson \&
Schwenn 2000; Gibson et al. 2006).

%The ideal magnetohydrodynamic (MHD) instabilities of the flux rope such
%as the kink instability and the torus instability are thought to be
%one type of mechanism that triggers the CME and associated
%activities (T{\"o}r{\"o}k \& Kliem 2003, 2005; Fan 2005; Kliem \&
%T{\"o}r{\"o}k 2006; Olmedo \& Zhang 2010). Thus, a detailed study of
%the flux rope is important for a clear understanding of CMEs, and
%this leads to a good ability to forecast CMEs and associated space
%weather.

The existence of flux ropes has been supported by using non linear
force-free field models based on observed vector magnetograms (Guo
et al. 2010; Canou \& Amari 2010; Jing et al. 2010). There are many
magnetohydrodynamic (MHD) simulations focused on the formation of
flux ropes and triggering mechanisms of eruption (Amari \& Luciani
1999; Aulanier et al. 2010; Fan \& Gibson 2004). Recently, the
multi-wavelength observations from the Atmospheric Imaging Assembly
(AIA; Lemen et al. 2012) onboard the \emph{Solar Dynamics
Observatory} (\emph{SDO}; Pesnell et al. 2012) provide us a good
opportunity to study the flux ropes. Direct observations of flux
ropes have been carried out by several authors since the launch of
\emph{SDO} (Cheng et al. 2011, 2012; Patsourakos et al. 2013; Li \&
Zhang 2013a, 2013b).

Since there exist homologous CMEs, it is natural to infer that if
there exist homologous flux ropes. Here, we define the homologous
flux ropes as follows: (1) the homologous flux ropes must originate
from the same region within the same active region (AR); (2) the
endpoints of the homologous flux ropes are anchored at the same
location; (3) the morphologies of the homologous flux ropes must
also resemble each other. In this letter, we present \emph{SDO}
observations of four homologous flux ropes and analyze their
evolution processes and the associated events.
%two events of flux ropes and investigate
%their fine-scale structures and magnetic properties by using the
%data from \emph{SDO}/AIA and the Helioseismic and Magnetic Imager
%(HMI; Schou \& Larson 2011).

\section{Observations and Data Analysis}

On 2013 May 20$-$22, four flux ropes were successively observed in
NOAA AR 11745. They are all above the neutral line of the AR and
their endpoints are rooted in the same region (see Figure 1). They
also have the generally similar shape and hence are homologous
according to their general resemblance in these aspects.

The \emph{SDO}/AIA takes full-disk images in 10 (E)UV channels at
1$\arcsec$.5 resolution and high cadence of 12 s. The flux ropes
could be clearly observed at two EUV channels of 131 {\AA} and 94
{\AA}. The 131 channel best shows the flux rope and we focus on this
channel in this study. We also present the 304 {\AA} observations in
order to analyze the relationship between the flux rope and the
associated filament. The 131 {\AA} channel corresponds to a high
temperature of about 11 MK (Fe VIII, Fe XXI) and the channel of 304
{\AA} (He II) is at 0.05 MK (O'Dwyer et al. 2010; Boerner et al.
2012; Parenti et al. 2012). We also use the full-disk line-of-sight
magnetic field data from the Helioseismic and Magnetic Imager (HMI;
Schou \& Larson 2011) onboard \emph{SDO}, with a cadence of $\sim$
45 s and a sampling of 0$\arcsec$.5 pixel$^{-1}$. The observations
from the Large Angle and Spectrometric Coronagraph (LASCO)
Experiment on board the Solar and Heliospheric Observatory (SOHO;
Brueckner et al. 1995) are also used to view the associated CME.

\section{Results}

\subsection{Overview of the Four Homologous Flux Ropes}

Figure 1 demonstrates the generally similar morphology and the
identical location of the four homologous flux ropes. The flux ropes
are composed of bright thread-like structures, which warp and
interweave together. The first three flux ropes are observed to
appear, rise up and then fade away in 1-2 hours. They are associated
with the activations and failed eruption of filaments. The fourth
flux rope is observed to erupt and then forms a CME with a velocity
of about 1200 km s$^{-1}$. It is associated with the partial
eruption of a filament and also an M5.0 flare.

\subsection{The First Three Flux Ropes}

The firstly observed flux rope occurred on 2013 May 20 and was
associated with the filament activation. At 09:24:10 UT, the middle
part of the filament was initially disturbed and associated with EUV
brightening (Figure 2(a) and Animation 1). Then the activated bright
filament material separately moved towards the two ends of the flux
rope. Meanwhile, the thread-like structures linking the south and
north filament material gradually became clear and the first flux
rope appeared (Figure 2(b)). The newly-formed bright arcades were
observed underneath the flux rope during the rise process of the
flux rope, which implies that the flare activity occurs at the
location of the arcades. At 10:15:10 UT, the flux rope developed to
its maximum and had the largest spatial scale and EUV intensity
(Figure 2(c)). Then the rise of the flux rope ceased and the EUV
intensity diminished little by little. At about 11:00:00 UT, the
flux rope ultimately faded away as it is difficult to be discerned
because of the low EUV intensity (Figure 2(d)). By comparing the 131
{\AA} observations with line-of-sight magnetograms, we find that the
northern end of the flux rope is rooted in positive polarity fields
and the southern end in negative polarity fields (Figure 2(c)).

At 09:59:10 UT on 2013 May 21, the middle part of the filament was
disturbed once again and the bright filament material gradually rose
up in the initial 10 min (Figures 3(a) and (e); Animations 2-131 and
2-304). Starting from 10:10:08 UT, the bright material moved toward
the north end of the filament. As seen in Figures 3(b) and (f), the
second flux rope could be only observed at 131 {\AA} and the
erupting filament material could be seen at both 131 {\AA} and 304
{\AA}. The erupting filament material seemed to be stranded and
frozen by the fine-scale structures of the flux rope, and rose up
with the rise of the entire flux rope. Similar to the first flux
rope, the new bright arcades were also observed below this flux rope
(Figure 3(c)), and associated with a C1.2 flare, which started at
10:23 UT, peaked at 10:40 UT and ended at 10:50 UT. The rise
distance of the flux rope was about 20 Mm and the average velocity
was approximately 30 km s$^{-1}$ between 10:30:10 UT and 10:40:10
UT. From 10:45:10 UT, the rise of the flux rope stopped and its EUV
intensity decreased subsequently. At 11:05:10 UT, the flux rope
became obscure and the erupting filament material seemed to fall
back to the solar surface, which implies the failed eruption of the
filament (Figures 3(d) and (h)).

Similar to the first two flux ropes, the activation of the filament
associated with the third flux rope started at the middle part at
01:12:10 UT on 2013 May 22 (Figure 2(e) and Animation 3). Then the
upper part of the filament lifted up and meanwhile some material
flow along the filament axis toward the north was observed clearly.
The third flux rope appeared at the location of the activated
filament at about 01:18:10 UT and afterwards rose up with the
filament material in it (Figure 2(f)). Starting from 01:24:10 UT,
the EUV intensity of the flux rope gradually decreased and the flux
rope was not detectable about 4 min later. Thereafter, at the
previous location of the flux rope, the material flow from the south
end to the north was observed from 02:01:08 UT. Then the third flux
rope appeared again (Figure 2(g)). At the rise process of the flux
rope, the flare arcades were formed below it (Figure 2(g)). These
arcades correspond to a C1.9 flare, which started at 02:25 UT,
peaked at 02:56 UT and ended at 03:08 UT. From 02:42:08 UT the
intensity of the flux rope started to decrease again and the flux
rope seemed to disappear at 03:00:08 UT (Figure 2(h)). The HMI
observations show that the northern end of the third flux rope is
rooted in positive polarity fields and the southern end in negative
polarity fields (Figure 2(g)), similar to the first two flux ropes.

\subsection{The Fourth Flux Rope and an Associated CME}

Before the appearance of the fourth flux rope, the middle part of
the filament was initially disturbed at 12:14:07 UT and associated
with EUV brightening, similar to the previous events. Then the
entire filament started to rise up with a speed of 45 km s$^{-1}$
and the helical structures were clearly observed (Figure 4(d);
Animations 4-131 and 4-304). The filament showed an obvious
anti-clockwise twist motion between 12:15:07 UT and 12:45:07 UT, and
the twist was estimated at 4$\pi$ (2 turns) by continuously tracking
the features of the helical structures according to AIA movies (see
Animation 4-304). In order to analyze the kinematic evolution of the
filament and the flux rope in detail, we obtain the stack plots
(Figures 5(a)$-$(b)) along slices ``A$-$B" and ``C$-$D" (Figures
4(b) and (e)). The erupting filament showed multiple spiral
structures and alternately dark and bright features in the stack
plots (Figure 5). This is caused by an interplay of two motions: the
overall uplift and the helical twist around the filament axis.

With the rise of the filament, the thread-like structures of the
fourth flux rope that seem to connect the north filament material
with its southern end started to appear at 12:23:03 UT in 131 {\AA}
observations (Figure 4(a)). As seen from Figure 5, the flux rope
initially rose slowly and the velocity was about 30 km s$^{-1}$.
Starting from 12:32:08 UT, the flux rope erupted rapidly and its
velocity increased to 130 km s$^{-1}$. It seems that the time of
12:32:08 UT is the turning point. Meanwhile, post-eruption arcades
appeared below the erupting flux rope at 131 {\AA} (Figure 4(b)).
However, these arcades could not be observed at low-temperature
channels such as 304 {\AA} and 171 {\AA} (Figures 4(e) and (f)). It
was also noticed that the flux rope and the filament were co-spatial
and showed consistent evolution (Figures 4(b) and (e)). At the late
phase, large part of the filament material fell back to the solar
surface (Figure 4(f) and Figure 5(b)). The sharpened image shows
that the flux rope is composed of multiple twisted structures, which
probably outline the magnetic field structures (Figure 4(c)).

At about 13:08:09 UT, an M5.0 flare occurred and the post-flare
loops at 304 {\AA} started to appear. The time interval between the
appearance of the post-eruption arcades observed at 304 {\AA} and
that at 131 {\AA} is about 36 min. The M5.0 flare increased to its
maximum at 13:32 UT and ended at 14:08 UT. The LASCO observations
showed that the eruption of the flux rope resulted in a fast CME
with an average speed of 1200 km s$^{-1}$ and a width angle of
182$\degr$ (Figures 4(g)$-$(h); Figures 5(c)$-$(d)). The kinematic
evolution of erupting flux rope at 131 {\AA} and bright twisted
structures (Figure 4(g)) of the CME is obtained (Figures
5(c)$-$(d)). The velocity of bright twisted structures of the CME
increased to 1800 km s$^{-1}$ at 13:42 UT and the corresponding
height was about 6.02 R$_{sun}$. Then the velocity decreased to 1100
km s$^{-1}$ at 15:06 UT. The profiles of the GOES soft X-ray 1$-$8
{\AA} flux and the CME velocity show similar trend. Based on the
kinematic evolution, it is deduced that the bright twisted
structures of the CME observed by LASCO correspond to the erupting
flux rope seen by AIA 131 {\AA}. We exclude the possibility that the
bright twisted structures correspond to the filament because very
little filament material heads out into space (Figures 4(e) and
5(b)).
%It seems like that the CME is composed of two parts: the leading
%edge and the bright structures of the flux rope. The twisted flux
%rope almost occupies the entire dark cavity, implying that the dark
%cavity corresponds to the erupting flux rope.

\section{Summary and Discussion}

We present the SDO/AIA observations of four homologous flux ropes on
2013 May 20$-$22 in AR 11745. The first three flux ropes gradually
rose up with a velocity of less than 30 km s$^{-1}$, and
subsequently their EUV intensities at 131 {\AA} decreased and the
flux ropes became obscure. \textbf{The fourth flux rope initially
underwent a ``slow-rise phase" with a velocity of 30$-$45 km
s$^{-1}$, and then accelerated rapidly to 130 km s$^{-1}$ and
erupted ultimately. It also resulted in a CME with a velocity of
about 1200 km s$^{-1}$.} The activated filament material was
spatially within the flux ropes and they showed consistent
evolution. Thus it is reasonable to substitute the kinematic
evolution of filament for that of the flux ropes in their early
stages. The observations definitely confirm the previous viewpoint
that the filament corresponds to the lower part of the flux rope,
where dense plasma collects.

Before the appearances of each flux rope, the brightenings along the
neutral line of the AR were observed from the upper photosphere to
the corona. Animation 1600 shows the evolution of the fourth flux
rope at the wavelength of 1600 {\AA}. The brightenings (especially
at 1600 {\AA}) probably implies the occurrence of magnetic
reconnection as the flux rope moves through the photosphere. Thus it
is deduced that the homologous flux ropes result from the continued
emergences of twisted flux ropes into the corona. As suggested by
Parker (1979), the field of a sunspot is composed of many separate
flux tubes, not a single large flux tube. The homologous flux ropes
probably originate from the same larger magnetic system that
consists of multiple smaller magnetic systems. The appearance of
each flux rope corresponds to the emergence of partial magnetic
structures of the larger magnetic system (Gibson et al. 2002;
Schrijver 2009). It is a pity that the evidence about the changes of
the magnetic field nearby the footpoints of the flux rope is not
notable because the AR is located near the solar limb. Okamoto et
al. (2008) and Kuckein et al. (2012) reported the brightenings in
H$\alpha$ images and concluded that the helical flux rope emerged
from below the photosphere based on the analysis of vector magnetic
field. Recently, the simulation results showed that the sustained
emergence of highly twisted magnetic fields resulted in repeated
formations of kink unstable flux ropes (Chatterjee \& Fan 2013).

The first three flux ropes were not associated with CMEs while the
fourth one erupted and led to a CME. The preceding events of flux
ropes may reduce the constraint of the fourth flux rope system by
the rearrangement of magnetic fields, which makes the fourth flux
rope easier to erupt. Animations 94 and STE-195 show the
observations of AIA 94 {\AA} and STEREO A 195 {\AA} on the day of
the eruption of the fourth flux rope. The overlying arcades above
the AR are partially removed at about 03:00 UT due to the third
event. Then an extended destabilization of overlying arcades
occurred at about 09:00 UT and resulted in the formation of
post-eruption arcades. Thus it is deduced that the preceding events
(especially the third event) probably have effects on the eruption
of fourth flux rope by removing the overlying arcades. The impact of
preceding eruption on the next eruption was widely investigated in
sympathetic eruptions (T{\"o}r{\"o}k et al. 2011; Schrijver et al.
2011; Bemporad et al. 2012). However, we could not completely
exclude the possibility that the eruption of the fourth flux rope is
an isolated event and not affected by the preceding events.

For the fourth flux rope, the associated filament showed an obvious
anti-clockwise twist motion at the initial stage, and the twist was
estimated at 4$\pi$. An magnetic flux rope becomes kink-unstable if
the twist exceeds a critical value of 2$\pi$ (Hood \& Priest 1981;
Fan 2005; T{\"o}r{\"o}k \& Kliem 2005). The amount of twisting in
this work is above the critical value, which indicates that kink
instability possibly triggers the early rise of the fourth flux
rope. Our observations are similar to the recent studies of Koleva
et al. (2012) and Kumar et al. (2012), who reported a larger twist
of about 6$\pi$$-$8$\pi$. It is also found that the post-eruption
arcades observed at 131 {\AA} appeared about 36 minutes earlier than
those at 304 {\AA} and 171 {\AA}. The delay implies that the plasma
of post-eruption arcades is cooling from high to low temperature
(Warren et al. 1999; Aschwanden \& Alexander 2001).

The homologous flux ropes could only be observed in hot channels
such as 94 and 131 {\AA}, which is consistent with recent
observations of Cheng et al. (2011) and Zhang et al. (2012).
However, some flux ropes are observed in all seven EUV channels
(304, 171, 193, 211, 335, 94, and 131 {\AA}) of the SDO/AIA (Li \&
Zhang 2013a, 2013b). Why do some flux rope have both the hot and
cool components, while others have only the hot component? It is
known that emerging ARs repeatedly produce flares and CMEs
(Schrijver 2009). However, the appearances and rises of the
homologous flux ropes occurred in the decaying phase of AR 11745.
How to understand this? The comprehensive characteristics of
homologous flux ropes need to be analyzed in further studies.

Our findings provide new clues for understanding the characteristics
of flux ropes. Firstly, there are multiple flux ropes that are
successively formed at the same location during an AR evolution
process. Secondly, a slow-rise flux rope does not necessarily result
in a CME, and a fast-eruption flux rope results in a CME. The
homologous flux ropes are not necessarily a series of homologous
CMEs, that is to say, the existence of homologous CMEs may need
stricter conditions. The cases of homologous flux ropes in the Sun
are probably more than homologous CMEs. We plan to thoroughly
examine the AIA data in future and deeply analyze the relationships
between homologous flux ropes and homologous CMEs.

\acknowledgments {We are grateful to Dr. B. Kliem for useful
discussions. We acknowledge the \emph{SDO}/AIA and HMI for providing
data. This work is supported by the National Basic Research Program
of China under grant 2011CB811403, the National Natural Science
Foundations of China (11303050, 11025315, 11221063 and 11003026) and
the CAS Project KJCX2-EW-T07.}

{}
\clearpage

\begin{figure}
\centering
\includegraphics
[bb=117 64 480 778,clip,angle=0,scale=0.7]{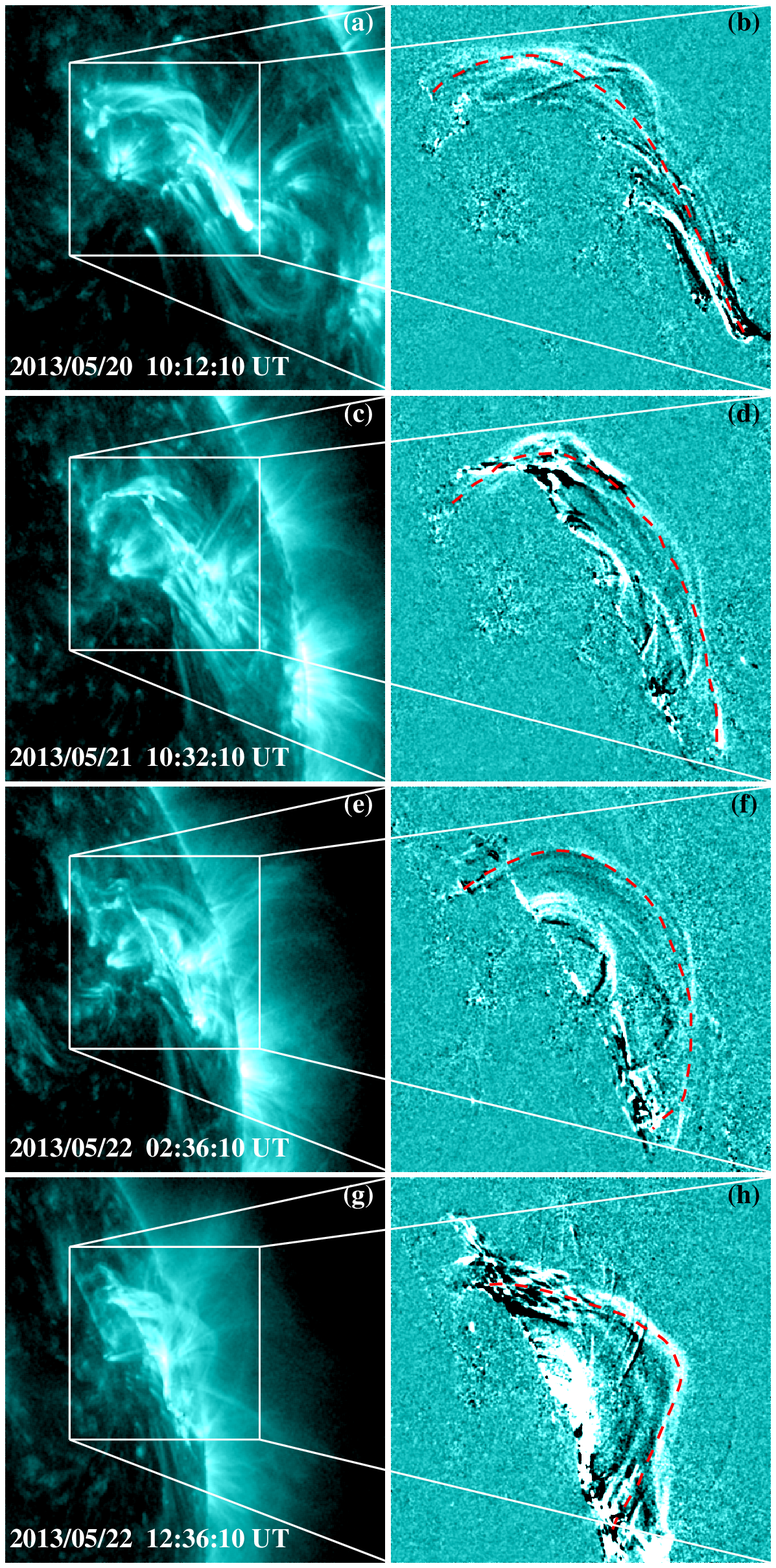}
\caption{Appearance of four homologous flux ropes in AR 11745 during
2013 May 20$-$22 as seen by {\emph{SDO}}$/$AIA 131 {\AA}
observations. The right column shows the difference images. The
white squares in left column denote the FOV of the images in right
column. Red dashed lines represent the main axes of the flux ropes.
\label{fig1}}
\end{figure}
\clearpage

\begin{figure}
\centering
\includegraphics
[bb=18 278 575 560,clip,angle=0,scale=0.9]{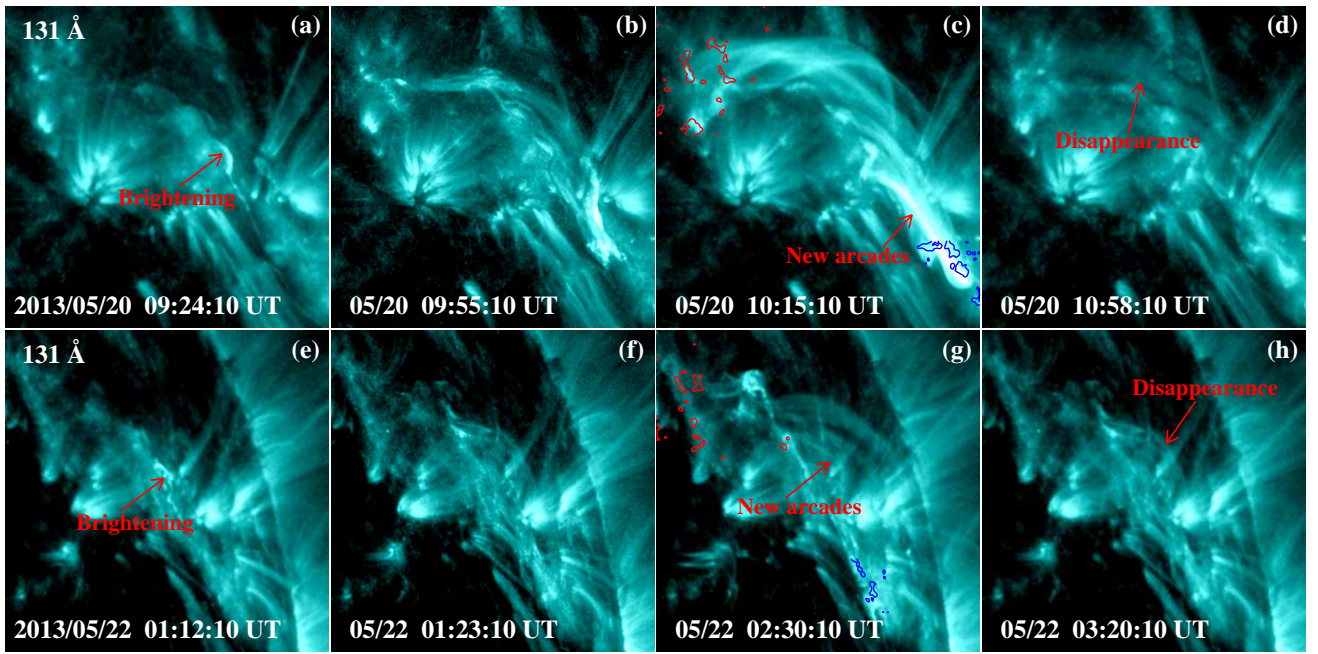}
\caption{Evolution of the first (panels (a)$-$(d)) and third (panels
(e)$-$(h)) flux ropes (see Animations 1 and 3, available in the
online edition of the journal). The red and blue contours in panels
(c) and (g) are the magnetic fields at $\pm$100 G levels at the
region of the endpoints of flux ropes. \label{fig2}}
\end{figure}
\clearpage

\begin{figure}
\centering
\includegraphics
[bb=18 278 577 562,clip,angle=0,scale=0.9]{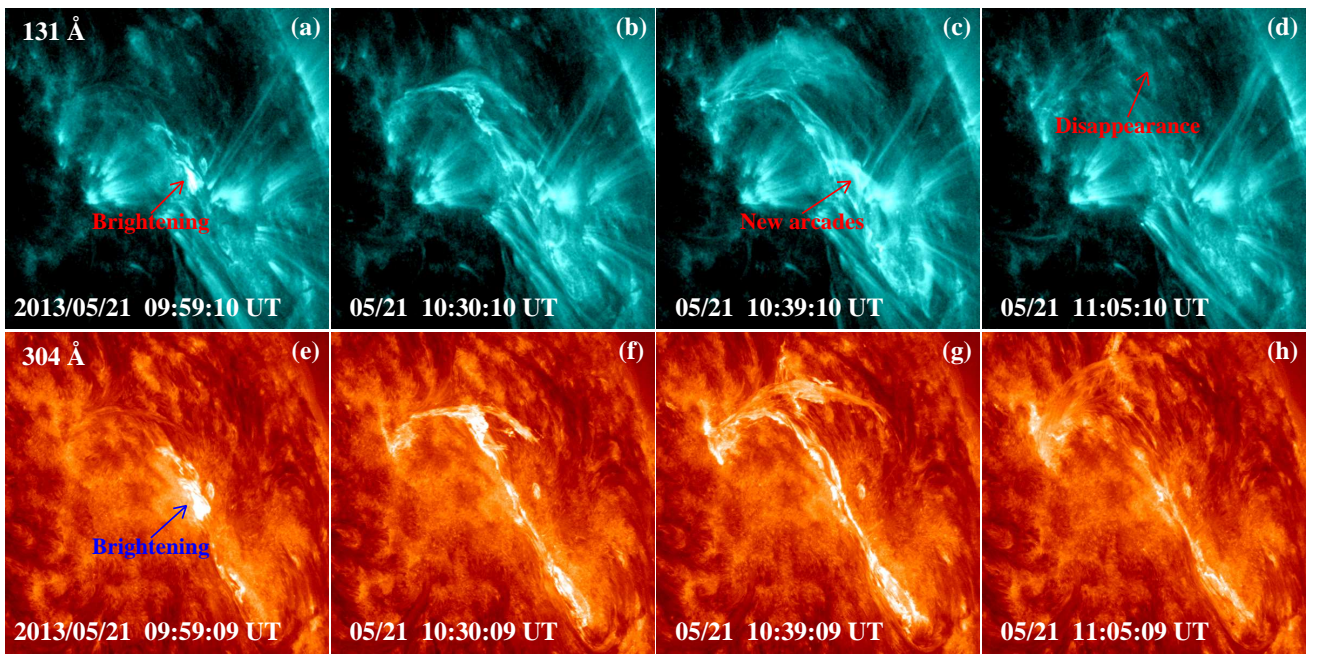}
\caption{Evolution of the second flux rope (panels (a)$-$(d)) and
the associated filament (panels (e)$-$(h); see Animations 2-131 and
2-304, available in the online edition of the journal).
\label{fig3}}
\end{figure}
\clearpage

\begin{figure}
\centering
\includegraphics
[bb=18 278 577 562,clip,angle=0,scale=0.9]{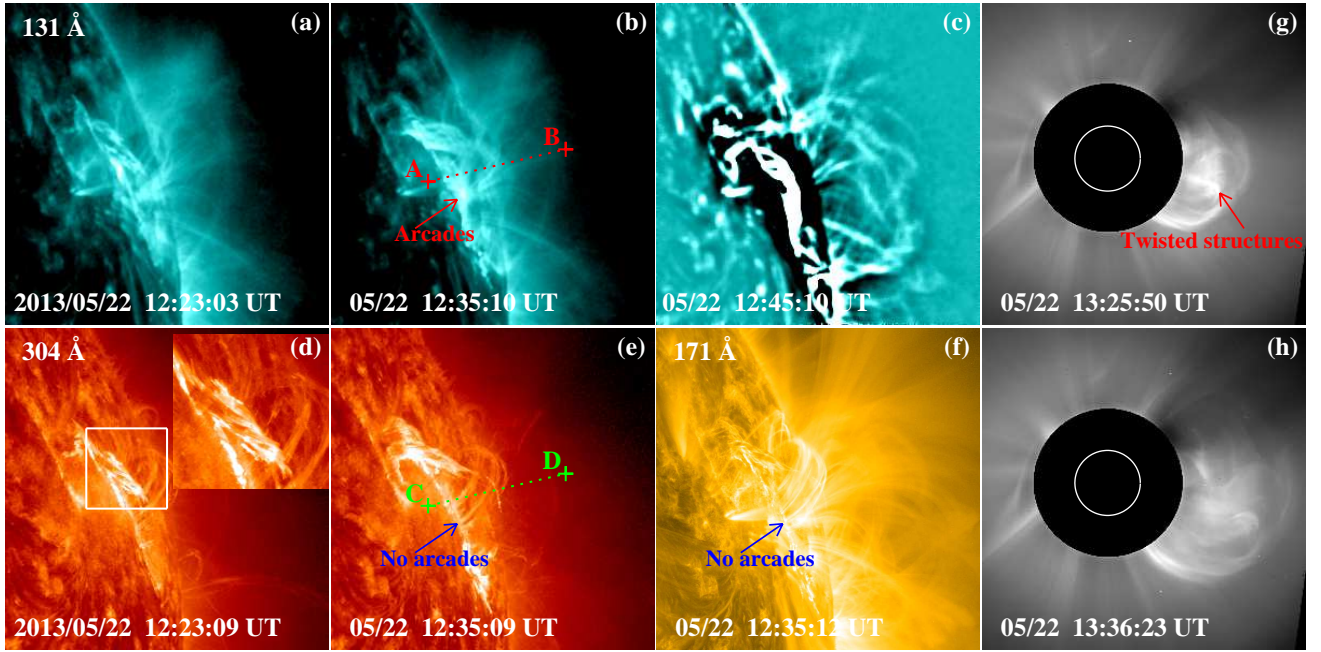}
\caption{Context information for the fourth flux rope and the
associated events. Panels (a)$-$(c): {\emph{SDO}}$/$AIA 131 {\AA}
images showing the eruption of the flux rope (see Animation 4-131);
panel (c) is the sharpened image by the process of unsharp masking.
Panels (d)$-$(e): {\emph{SDO}}$/$AIA 304 {\AA} images showing the
filament eruption (see Animation 4-304); white square denotes the
FOV of the small image in the top right corner of panel (d). Panel
(f): {\emph{SDO}}$/$AIA 171 {\AA} image showing no post-eruption
arcades (compared with 131 {\AA} in panel (b)). Panels (g)$-$(h):
LASCO C2 images showing the CME associated with the eruption of the
flux rope.}
\end{figure}
\clearpage

\begin{figure}
\centering
\includegraphics
[bb=94 218 506 644,clip,angle=0,scale=0.9]{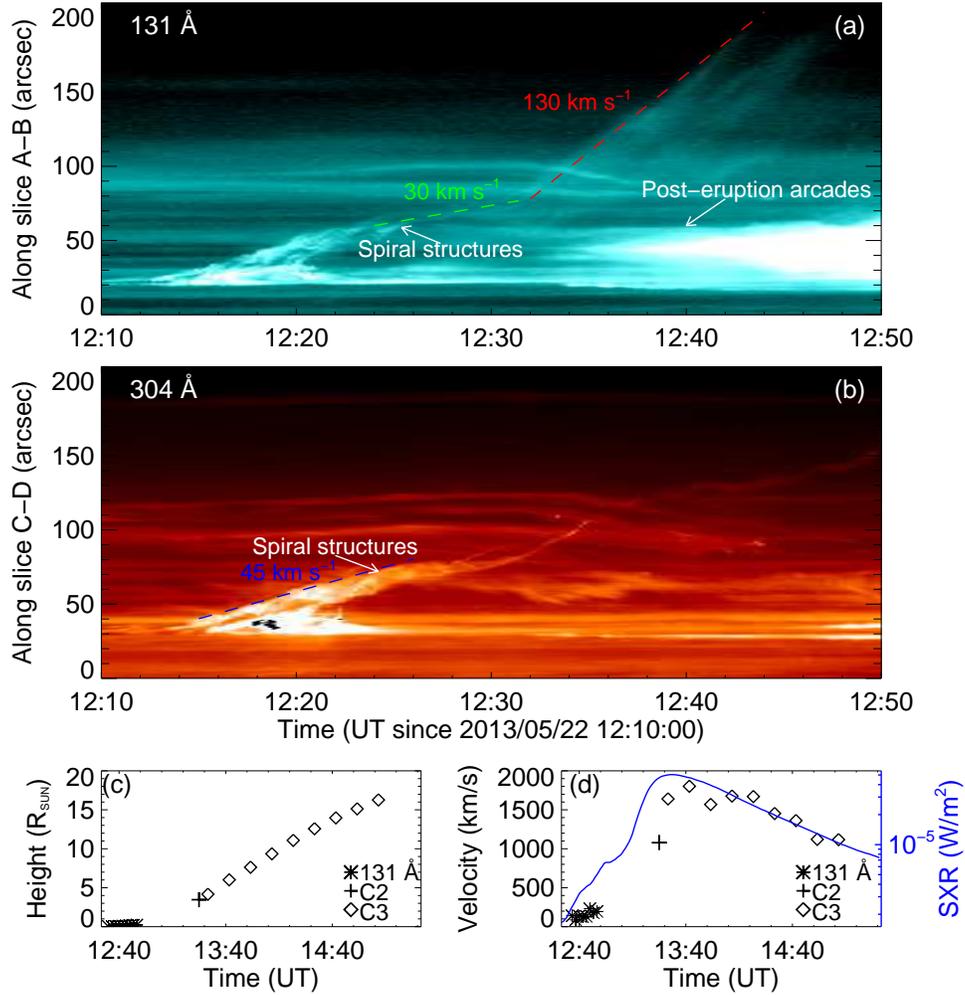} \caption{Panels
(a)$-$(b): stack plots along slices ``A$-$B" (red dashed line in
Figure 4(b)) and ``C$-$D" (green dashed line in Figure 4(e)) showing
the evolution of the flux rope and the filament. Panels (c)$-$(d):
height-time and velocity-time profiles of erupting flux rope at AIA
131 {\AA} and bright twisted structures observed by LASCO/C2 and C3;
the blue curve in panel (d) denotes the GOES SXR 1$-$8 {\AA} flux of
the associated flare. \label{fig5}}
\end{figure}
\clearpage

\end{document}